# The principle of stationary action in neural systems

Erik D. Fagerholm[1], Karl J. Friston[2], Rosalyn J. Moran[1], Robert Leech[1]

[1] Department of Neuroimaging, King's College London
[2] Wellcome Centre for Human Neuroimaging, University College London

Corresponding author: erik.fagerholm@kcl.ac.uk

**Abstract**

The principle of stationary action is a cornerstone of modern physics, providing a powerful framework for investigating dynamical systems found in classical mechanics through to quantum field theory. However, computational neuroscience, despite its heavy reliance on concepts in physics, is anomalous in this regard as its main equations of motion are not compatible with a Lagrangian formulation and hence with the principle of stationary action. Taking the Dynamic Causal Modelling neuronal state equation, Hodgkin-Huxley model, and the Leaky Integrate-and-Fire model as examples, we show that it is possible to write complex oscillatory forms of these equations in terms of a single Lagrangian. We therefore bring mathematical descriptions in computational neuroscience under the remit of the principle of stationary action and use this reformulation to express symmetries and associated conservation laws arising in neural systems.

**Introduction**

In classical physics, dynamical systems are framed in terms of equations of motion describing quantities such as position, velocity, and acceleration. On the other hand, the equations of motion used in computational neuroscience refer to more abstract quantities, such as membrane potentials, firing rates, and macroscopic neuronal activity (Breakspear, 2017). However, these formulations have thus far been expressed in mathematical frameworks that are incompatible with the principle of stationary action (Deco et al., 2008). Given the practical advantages of applying variational principles, our goal is to take the first step towards reformulating computational neuroscience in terms of the principle of stationary action. To do so, we will show that (under certain conditions) it is possible to express modified versions of three broadly used dynamics in neuroscience in terms of a single Lagrangian. These dynamics are given by: a) the Dynamic Causal Modelling (DCM) neuronal state equation (Friston et al., 2003); b) the Hodgkin-Huxley (HH) equations (Hodgkin and Huxley, 1952); and c) the Leaky Integrate-and-Fire (LIF) model (Abbott, 1999). We use these examples to demonstrate that the full range of dynamics displayed by the three formulations can be unified in terms of a single mathematical framework that is compatible with the principle of stationary action.

Most problems in classical physics are phrased in terms of a common question – given initial conditions, how will a dynamical system evolve? These problems are addressed by analysing equations of motion, which can be likened to a machinery that receives the system's current state as an input and produces an output, showing the state at an infinitesimally small step ahead in time. Let us, for instance, consider the trajectory of a cannonball in mid-flight. At any point in its path, the cannonball's position $r(t)$ can be calculated via Newton's second law in a gravitational potential:

$$\frac{d^2[r(t)]}{dt^2} = -\frac{GM}{r^2(t)} \qquad [1]$$

where $G$ is the universal gravitational constant and $M$ is the mass of the earth.

If we were to simulate the path of the cannonball, we would begin by transforming Eq [1] into two first order differential equations:

$$\frac{dr}{dt} = u$$

$$\frac{du}{dt} = -\frac{GM}{r^2}$$

[2]

which can then be transformed to a pair of difference equations as follows:

$$r_{t+1} = r_t + u_t \Delta t$$

$$u_{t+1} = u_t - \frac{GM}{r_t^2} \Delta t$$

[3]

i.e. at every iteration of the numerical integration, we enter the position of the cannonball at the present timepoint $r_t$ and obtain its position one small step in time $\Delta t$ in the future $r_{t+1}$.

However, there is another way in which we can calculate the path of the cannonball, without needing to calculate its trajectory in such incremental steps. This alternative method is known as the principle of stationary action and in order for it to be employed we must begin by casting the relevant equation of motion in the form of a Lagrangian $\mathcal{L}$. The latter is a function of the system's state and the rate of change of its state, which—in the example of the cannonball— means its position and velocity: $\mathcal{L} = \mathcal{L}(r, \dot{r})$. In classical mechanics, the Lagrangian is usually defined as the difference between the system's kinetic $T$ and potential $U$ energy which, for a projectile moving in a gravitational potential, takes the following form:

$$\mathcal{L} = T - U = \frac{1}{2}m\dot{r}^2 + \frac{GmM}{r}$$

[4]

where $m$ is the mass of the cannonball.

We now turn to the action $S$, which is a functional of the Lagrangian defined as follows between an initial $t_i$ and final $t_f$ timepoint:

$$S = \int_{t_i}^{t_f} \mathcal{L} dt$$

[5]

The principle of stationary action (or Hamilton's principle) then tells us that a small variation of any path that satisfies the underlying equation of motion leaves the action unchanged to within a first-order approximation:

$$\delta S = 0. \qquad [6]$$

Standard arguments (Morse and Feshbach, 1953) can then be followed to show that this constraint implies that the equation of motion can be recovered by using the Euler-Lagrange equation:

$$\frac{d}{dt}\left[\frac{\partial \mathcal{L}}{\partial \dot{r}}\right] = \frac{\partial \mathcal{L}}{\partial r}. \qquad [7]$$

We can verify that this is correct for Eq [4] by calculating the left-hand side of Eq [7]:

$$\frac{d}{dt}\left[\frac{\partial \mathcal{L}}{\partial \dot{r}}\right] = \frac{d}{dt}[m\dot{r}] = m\ddot{r} \qquad [8]$$

and similarly, for the right-hand side of Eq [7]:

$$\frac{\partial \mathcal{L}}{\partial r} = -\frac{GmM}{r^2} \qquad [9]$$

Using Eqs [7] through [9] we see that the original equation of motion [1] is indeed recovered via the principle of stationary action.

Virtually all of modern physics has been formulated in terms of the principle of stationary action, from Maxwell's equations in electromagnetism (Landau, 2013), to the Dirac equation in quantum mechanics (Shankar, 2012). This approach has yielded many advantages. Firstly, disparate equations of motion can be described in terms of a single Lagrangian function, allowing for a parsimonious unified mathematical framework. This was famously demonstrated via the single Lagrangian formulation of the entire standard model of particle physics (Cottingham and Greenwood, 2007). Secondly, the path integral formulation of the principle of stationary action allows for the incremental trajectory calculation in Eq [3] to be replaced with a single computational step. In other words, given that we know the initial and final states of the system, as well as how much time elapses, we can – in a single step – draw the path that will be followed by the cannonball; namely, the path that renders the action stationary.

Every other path would increase the value of the action[1]. Thirdly, and perhaps of most interest in the study of the brain, is that mechanical similarity (Landau, 1976) can only be quantified in terms of a Lagrangian formulation. Mechanical similarity entails a system that continues to follow a single equation of motion following a space/time scale transformation. In such a system, a scaled Lagrangian $\mathcal{L}_{scaled}$ differs only from the original (unscaled) Lagrangian $\mathcal{L}$ by a multiplicative constant $k$, such that:

$$\mathcal{L}_{scaled} = k\mathcal{L} \qquad [10]$$

Applying Eq [8] to the scaled Lagrangian we obtain:

$$\frac{d}{dt}\left[\frac{\partial \mathcal{L}_{scaled}}{\partial \dot{r}}\right] = \frac{\partial \mathcal{L}_{scaled}}{\partial r} \qquad [11]$$

which, using Eq [10], reads:

$$\frac{d}{dt}\left[\frac{\partial (k\mathcal{L})}{\partial \dot{r}}\right] = \frac{\partial (k\mathcal{L})}{\partial r} \qquad [12]$$

or equivalently:

$$k\frac{d}{dt}\left[\frac{\partial \mathcal{L}}{\partial \dot{r}}\right] = k\frac{\partial \mathcal{L}}{\partial r} \qquad [13]$$

in which the constant $k$ cancels, thus leaving the equation of motion for the original (unscaled) system. This highlights an important point – information that can be extracted from the scale transformation is only accessible if the Lagrangian equations of motion are available. This can be seen by the fact that the value of the constant $k$ appears in the Lagrangian formulation in Eq [10], but is 'invisible' in the non-Lagrangian formulation in Eq [13]. This is particularly relevant to neural systems, where we would like to characterise the scaling behaviour of functional architectures, both across the lifespan of an organism as well as across different species (Buzsaki et al., 2013). Finally, Lagrangian formulations allow for otherwise unattainable insight into conservation laws that arise by virtue of symmetries in the mathematical expression of a physical law, where a symmetry is any transformation that leaves the action unchanged. Specifically, Noether's theorem (Noether, 1918) tells us that for every symmetry there is an associated conserved quantity $N$ given by:

$$N = \left(\frac{\partial \mathcal{L}}{\partial \dot{r}}\dot{r} - \mathcal{L}\right)\delta t - \frac{\partial \mathcal{L}}{\partial \dot{r}}\delta r = \mathcal{H}\delta t - p\delta r \qquad [14]$$

where $\mathcal{H}$ is the system's Hamiltonian (total energy) and $p$ is its momentum.

---

[1] Or decrease it, depending upon whether the path minimises or maximises action. This is why we use the more correct term principle of 'stationary' action, as opposed to principle of 'least' action.

We see, therefore, that the expression for the conserved quantities in Eq [14] requires the system's equation of motion to be cast in the form of a Lagrangian $\mathcal{L}$. Given these insights into the mechanics of dynamical systems – which are precluded in a non-Lagrangian framework – it would clearly be beneficial for the equations of motion in computational neuroscience to be written in the language of Lagrangian mechanics. This is the spirit in which we proceed, by showing that it is possible to cast three broadly used equations – the DCM neuronal state equation, HH model, and LIF model – in the form of a single Lagrangian function.

**Methods & Results**

**The DCM neuronal state equation:** A generic non-linear dynamical system can be expressed in terms of a Taylor series expansion (Stephan et al., 2008), which in its simplest form, for the $i^{th}$ region, is expressed as the linear DCM neuronal state equation:

$$\dot{z}_i(t) = \sum_j A_j z_j(t) + \sum_j C_j v_j(t) + \omega_i^{(x)}(t), \qquad [15]$$

where $z$ represents the state of the system in question; $A$ is the intrinsic coupling matrix; $C$ is the extrinsic input matrix; $v = u + \omega^{(v)}$, where $\omega^{(v)}$ is a noise term describing random, non-Markovian fluctuations on external perturbations $u$; and $\omega^{(x)}$ is a noise term describing random, non-Markovian fluctuations on $z$ (Li et al., 2011). Using Eq [15], we can describe arbitrary dynamical systems in a way that allows for Bayesian model inversion, which allows us to obtain estimates of latent model parameters in the presence of noise on states. These model parameters include the ways in which the system in question is connected, both intrinsically as well as extrinsically with the surrounding environment.

**The HH model:** The three equations describing the potassium channel activation $n$, sodium channel activation $m$, and sodium channel inactivation $h$ within the HH model are written as follows:

$$\dot{n} = \alpha_n(V_m)(1-n) - \beta_n(V_m)n$$
$$\dot{m} = \alpha_m(V_m)(1-m) - \beta_m(V_m)m \quad [16]$$
$$\dot{h} = \alpha_h(V_m)(1-h) - \beta_h(V_m)h$$

where $\alpha$ and $\beta$ are rate constants and $V_m$ is the membrane potential.

**The LIF model:** The LIF model describes action potentials produced at the individual cell level in terms of a current $I(t)$ that produces a potential $V(t)$ by charging a capacitor:

$$\tau \frac{dV}{dt} = -(V - V_L) + RI \quad [17]$$

where $\tau$ is a time constant defined as $\tau = RC$; $R$ is the resistance; $C$ is the capacitance; and $V_L$ is the leak (resting) potential in the absence of external perturbation.

**A failed attempt using Real variables:** Using the DCM equation [15] as an example, we will now show why complex variables are necessary when casting any equation of motion that is first order in time in the form of a Lagrangian. We will prove this first by contradiction: i.e. we begin by attempting to use Real variables and demonstrate that this leads to a non-unique solution. If we insist on using Real variables then (as shown below) we must assume that $v$ and $\omega$ are constant in time, in which case we can re-write Eq [15] as follows:

$$\dot{z}_i(t) = \sum_j A_j x_j(t) + d_i, \quad [18]$$

where $d_i = \sum_j C_j v_j + \omega_i^{(x)}$ is a constant.

Differentiating Eq [18] with respect to time we obtain:

$$\ddot{z}_i(t) = \sum_j A_j \dot{z}_j(t), \quad [19]$$

which, together with Eq [18], gives:

$$\ddot{z}_i(t) = \sum_j E_j z_j(t) + f_i,$$ [20]

where $E = A^2$ and $f_i = \sum_j A_j d_j$.

We then note that, as long as the $E$ matrix is symmetric, Eq [20] can be viewed as the Euler-Lagrange equation:

$$\frac{d}{dt}\left[\frac{\partial \mathcal{L}}{\partial \dot{z}_i}\right] = \frac{\partial \mathcal{L}}{\partial z_i},$$ [21]

for the Lagrangian given by:

$$\mathcal{L} = \frac{1}{2}\sum_j \left(\dot{z}_j^2(t) + f z_j(t)\right) + \sum_{j,k} E_{j,k} z_j(t) z_k(t),$$ [22]

thereby giving a Lagrangian description of Eq [20].

However, the problem with this approach is that, although any solution of the original DCM with time-independent driving inputs and noise in Eq [18] is also a solution of the second-order Euler-Lagrange equation [21], the reverse is not true: i.e., the second order Euler-Lagrange equation also has other solutions that are not valid solutions of the original DCM recovery model. Therefore, if we insist upon using real variables, it is not possible to cast the first-order DCM in the form of a Lagrangian in a way that allows for time-dependent external inputs and noise, or for the DCM neuronal state equation to be uniquely recovered via the Euler-Lagrange equation.

**The complex oscillatory equations of motion:** We will now demonstrate that it is possible to cast the DCM neuronal state equation [15], the HH model in Eq [16], and the LIF model in Eq [17] in the form of a single Lagrangian as long as the following three conditions are met: a) all dependent variable are complex; b) the $A$ matrix in Eq [15] is Hermitian; and c) we can introduce an imaginary unit $i$ on the left-hand sides of the equations in all three models – thus rendering the solutions oscillatory (see Discussion).

We write the DCM neuronal state equation [15] modified according to conditions a), b) and c) as follows:

$$i\dot{z}_i = \sum_j A_j z_j + \sum_j C_j v_j + \omega_i^{(z)}, \quad [23]$$

Similarly, we write the HH model in Eq [16] modified according to conditions a) and c) as follows:

$$i\dot{n} = \alpha_n(V_m)(1-n) - \beta_n(V_m)n$$
$$i\dot{m} = \alpha_m(V_m)(1-m) - \beta_m(V_m)m \quad [24]$$
$$i\dot{h} = \alpha_h(V_m)(1-h) - \beta_h(V_m)h$$

Finally, we write the LIF model in Eq [17] modified according to a) and c) as follows:

$$i\tau \frac{dV}{dt} = -(V - V_L) + RI \quad [25]$$

**The complex oscillatory DCM-HH-LIF Lagrangian**: We are now in a position to make the central proposition of this paper, i.e. that Eqs [23] through [25] can be cast in the form of the following single Lagrangian:

$$\mathcal{L} = \frac{i}{2}\left(\sum_j (z_j^* \dot{z}_j - z_j \dot{z}_j^*) + n^*\dot{n} - n\dot{n}^* + m^*\dot{m} - m\dot{m}^* + h^*\dot{h} - h\dot{h}^* + V^*\dot{V} - V\dot{V}^*\right)$$
$$- \sum_{j,k} z_k^* A_{jk} z_j - \sum_j \left(\sum_k C_k v_k + \omega_j^{(z)}\right)(z_j + z_j^*) + \beta_n(V_m)nn^*$$
$$+ \alpha_n(V_m)(nn^* - n - n^*) + \beta_m(V_m)mm^* \quad [26]$$
$$+ \alpha_m(V_m)(mm^* - m - m^*) + \beta_h(V_m)hh^*$$
$$+ \alpha_h(V_m)(hh^* - h - h^*) + \frac{1}{\tau}VV^* - \frac{1}{\tau}(RI + V_L)(V + V^*),$$

where we use star notation to indicate complex conjugation. We will now show that this proposed Lagrangian is correct by verifying that each of the three equations of motion [23] through [25] can be recovered from Eq [26] via the principle of stationary action.

**Recovering the complex oscillatory DCM neuronal state equation:** In order to show that Eq [26] is of the correct form, we temporarily proceed under the assumption that the variables, e.g. $z$ and $z^*$ are not related by complex conjugation. We therefore write the general variation of $\mathcal{L}(z, z^*)$, which we treat as a function of two independent variables $(z, z^*)$, as follows:

$$d\mathcal{L} = \frac{\partial \mathcal{L}}{\partial z} dz + \frac{\partial \mathcal{L}}{\partial z^*} dz^*. \qquad [27]$$

Since the values of $z$ and $z^*$ and their variations $dz$ and $dz^*$ are all arbitrary, this equation must also hold when we add restrictions, e.g. by insisting that $z^*$ must be equal to the complex conjugate of $z$ and that $dz^*$ must be equal to the complex conjugate of $dz$. In other words, although we begin by treating $z$ and $z^*$ as independent variables, Eq [27] must also hold when $z$ and $z^*$ are related by complex conjugation.

Next, we consider two specific variations:

$$\begin{aligned} dz = du + idv & \quad \rightarrow \quad dz^* = du - idv, \\ dz = du - idv & \quad \rightarrow \quad dz^* = du + idv, \end{aligned} \qquad [28]$$

where $du$ and $dv$ are arbitrary real variations.

If we now look for stationary points of $\mathcal{L}$, then $d\mathcal{L}$ must be zero for both of the above variations in Eq [28], implying that:

$$\begin{aligned} d\mathcal{L} = 0 = \frac{\partial \mathcal{L}}{\partial z}(du + idv) + \frac{\partial \mathcal{L}}{\partial z^*}(du - idv), \\ d\mathcal{L} = 0 = \frac{\partial \mathcal{L}}{\partial z}(du - idv) + \frac{\partial \mathcal{L}}{\partial z^*}(du + idv), \end{aligned} \qquad [29]$$

from which we see that:

$$\begin{aligned} 0 = \left(\frac{\partial \mathcal{L}}{\partial z} + \frac{\partial \mathcal{L}}{\partial z^*}\right) du = 0, \\ 0 = \left(\frac{\partial \mathcal{L}}{\partial z} - \frac{\partial \mathcal{L}}{\partial z^*}\right) i dv = 0, \end{aligned} \qquad [30]$$

and since $du$ and $dv$ are arbitrary, this implies that:

$$\frac{\partial \mathcal{L}}{\partial z} = \frac{\partial \mathcal{L}}{\partial z^*} = 0. \qquad [31]$$

Note that both conditions in Eq [31] must hold, i.e. if $\mathcal{L}$ is Real (when $z^*$ is the complex conjugate of $z$), then $\frac{\partial \mathcal{L}}{\partial z}$ and $\frac{\partial \mathcal{L}}{\partial z^*}$ are complex conjugates of each other and the two conditions reduce to one.

In light of the above, let us now take partial derivatives of our proposed Lagrangian in Eq [26] with respect to both $z$ and its complex conjugate $z^*$:

$$\frac{\partial \mathcal{L}}{\partial z_i^*} = \frac{i}{2}\dot{z}_i - \omega_i^{(z)} - \sum_j A_j z_j - \sum_j C_j v_j,$$

$$\frac{\partial \mathcal{L}}{\partial z_i} = -\frac{i}{2}\dot{z}_i^* - \omega_i^{(z)} - \sum_j z_j^* A_j - \sum_j C_j v_j \qquad [32]$$

These two equations are complex conjugates of each other as long as the $A$ matrix is Hermitian, i.e. whenever one equation is true, so is the other and we therefore do not need to solve them separately. Note that it is for this reason that we must include an imaginary unit $i$ in the Lagrangian formulation – this addition flips the signs of the $\frac{\partial}{\partial t}$ terms in the complex conjugation of the Lagrangian in Eq [26] and hence makes the derivatives with respect to $z$ and $z^*$ complex conjugates of one another. We can also use integration by parts in time to show that (dropping the subscript):

$$\int_{-\infty}^{+\infty} dt\, z^* i\frac{dz}{dt} = \int dt \left(i\frac{dz}{dt}\right)^* z, \qquad [33]$$

as long as $z(t)$ tends to zero rapidly as $t$ tends to $+\infty$ or $-\infty$. In other words, $i\frac{d}{dt}$ is a Hermitian operator in the time domain.

We now work out the other term of the Euler-Lagrange equation from Eq [26]:

$$\frac{d}{dt}\left[\frac{\partial \mathcal{L}}{\partial \dot{z}_i^*}\right] = -\frac{i}{2}\dot{z}_i,$$

$$\frac{d}{dt}\left[\frac{\partial \mathcal{L}}{\partial \dot{z}_i}\right] = \frac{i}{2}\dot{z}_i^*,$$

[34]

which, using the Euler-Lagrange equations:

$$\frac{\partial \mathcal{L}}{\partial z_i^*} = \frac{d}{dt}\left[\frac{\partial \mathcal{L}}{\partial \dot{z}_i^*}\right],$$

$$\frac{\partial \mathcal{L}}{\partial z_i} = \frac{d}{dt}\left[\frac{\partial \mathcal{L}}{\partial \dot{z}_i}\right],$$

[35]

together with Eq [32] gives:

$$i\dot{z}_i = \sum_j A_j z_j + \sum_j C_j v_j + \omega_i^{(z)},$$

$$-i\dot{z}_i^* = \sum_j z_j^* A_j + \sum_j C_j v_j + \omega_i^{(z)},$$

[36]

i.e. we recover the complex oscillatory DCM neuronal state equation [23] and its adjoint. Note that, unlike the example case using real variables in Eq [18] through Eq [22], the use of complex variables allows for the modified state equation [23] to be uniquely recovered in a way that allows for both time-dependent external inputs and noise terms.

**Recovering the complex oscillatory HH model;** Following the same logic presented in Eq [27] through Eq [36], we now take the following variations with respect to Eq [26], taking the variable $n$ as an example:

$$\frac{\partial \mathcal{L}}{\partial n^*} = \frac{i}{2}\dot{n} + \beta_n(V_m)n - \alpha_n(V_m)(1-n)$$

$$\frac{\partial \mathcal{L}}{\partial n} = -\frac{i}{2}\dot{n}^* + \beta_n(V_m)n^* - \alpha_n(V_m)(1-n^*)$$

[37]

which, together with:

$$\frac{d}{dt}\left[\frac{\partial \mathcal{L}}{\partial \dot{n}^*}\right] = -\frac{i}{2}\dot{n}$$

$$\frac{d}{dt}\left[\frac{\partial \mathcal{L}}{\partial \dot{n}}\right] = \frac{i}{2}\dot{n}^*$$

[38]

and the Euler-Lagrange equations:

$$\frac{d}{dt}\left[\frac{\partial \mathcal{L}}{\partial \dot{n}^*}\right] = \frac{\partial \mathcal{L}}{\partial n^*}$$

$$\frac{d}{dt}\left[\frac{\partial \mathcal{L}}{\partial \dot{n}}\right] = \frac{\partial \mathcal{L}}{\partial n}$$

[39]

gives us the following:

$$i\dot{n} = \alpha_n(V_m)(1-n) - \beta_n(V_m)n$$

$$-i\dot{n}^* = \alpha_n(V_m)(1-n^*) - \beta_n(V_m)n^*$$

[40]

i.e. upon repeating the steps shown in Eqs [37] through [40], we recover the complex oscillatory Hodgkin-Huxley equations in Eq [16] and their adjoints.

**Recovering the complex oscillatory LIF model;** Following the same logic presented in Eq [27] through Eq [36], we now take the following variations with respect to Eq [26]:

$$\frac{\partial \mathcal{L}}{\partial V^*} = \frac{i}{2}\dot{V} + \frac{1}{\tau}V - \frac{1}{\tau}(RI + V_L)$$

$$\frac{\partial \mathcal{L}}{\partial V} = -\frac{i}{2}\dot{V}^* + \frac{1}{\tau}V^* - \frac{1}{\tau}(RI + V_L)$$

[41]

which, together with:

$$\frac{d}{dt}\left[\frac{\partial \mathcal{L}}{\partial \dot{V}^*}\right] = -\frac{i}{2}\dot{V}$$

$$\frac{d}{dt}\left[\frac{\partial \mathcal{L}}{\partial \dot{V}}\right] = \frac{i}{2}\dot{V}^*$$

[42]

and the Euler-Lagrange equations:

$$\frac{d}{dt}\left[\frac{\partial \mathcal{L}}{\partial \dot{V}^*}\right] = \frac{\partial \mathcal{L}}{\partial V^*}$$

$$\frac{d}{dt}\left[\frac{\partial \mathcal{L}}{\partial \dot{V}}\right] = \frac{\partial \mathcal{L}}{\partial V}$$

[43]

gives us the following:

$$i\tau \dot{V} = -(V - V_L) + RI$$
$$-i\tau \dot{V}^* = -(V - V_L) + RI \qquad [44]$$

i.e. we recover the complex oscillatory LIF equation [25] and its adjoint. We have therefore verified the Lagrangian in [26], as it correctly encodes the complex oscillatory forms of the DCM neuronal state equation [23], the HH model in Eq [24], and the LIF equation [25].

**Neural momentum:** Using Noether's theorem in Eq [14], we can use Eq [26] to write an expression for the canonical momentum conjugate to the $i^{th}$ neural region – e.g. for the $z$ variable in Eq [23]:

$$p_{z_i} = \frac{\partial \mathcal{L}}{\partial \dot{z}_i} = \frac{i}{2} z_i^* \qquad [45]$$

Note, however, that this quantity is not conserved, as none of the three models considered here are translationally invariant.

**Neural Energy:** Again using Noether's theorem in Eq [14], we can use Eq [26] to write an expression for the quantity that is conserved by virtue of time translational invariance, i.e. the system's total energy, or Hamiltonian $\mathcal{H}$:

$$\begin{aligned}\mathcal{H} &= \frac{\partial \mathcal{L}}{\partial \dot{z}}z + \frac{\partial \mathcal{L}}{\partial \dot{z}^*}z^* + \frac{\partial \mathcal{L}}{\partial \dot{n}}n + \frac{\partial \mathcal{L}}{\partial \dot{n}^*}n^* + \frac{\partial \mathcal{L}}{\partial \dot{m}}m + \frac{\partial \mathcal{L}}{\partial \dot{m}^*}m^* + \frac{\partial \mathcal{L}}{\partial \dot{h}}h + \frac{\partial \mathcal{L}}{\partial \dot{h}^*}h^* + \frac{\partial \mathcal{L}}{\partial \dot{V}}V + \frac{\partial \mathcal{L}}{\partial \dot{V}^*}V^* - \mathcal{L} \\ &= \sum_{j,k} z_k^* A_{jk} z_j + \sum_j \left(\sum_k C_k v_k + \omega_j^{(z)}\right)(z_j + z_j^*) - \beta_n(V_m)nn^* \\ &\quad + \alpha_n(V_m)(n + n^* - nn^*) - \beta_m(V_m)mm^* + \alpha_m(V_m)(m + m^* - mm^*) \\ &\quad - \beta_h(V_m)hh^* + \alpha_h(V_m)(h + h^* - hh^*) - \frac{1}{\tau}VV^* + \frac{1}{\tau}(RI + V_L)(V + V^*),\end{aligned} \qquad [46]$$

Note, however, that this system is dissipative as it is influenced by external perturbations $v$ and $\omega$ and time-dependent current $I$ and hence its energy is not conserved due to the lack of time translational invariance in the DCM neuronal state equation. On the other hand, if we consider non-dissipative versions of the DCM neuronal state equation and the LIF model, i.e.

where $v$, $\omega$, and $I$ are constant (zero for convenience), then we can use Eq [26] to show that the new Lagrangian reads:

$$\mathcal{L} = \frac{i}{2}\left(\sum_j\left(z_j^*\dot{z}_j - z_j\dot{z}_j^*\right) + n^*\dot{n} - n\dot{n}^* + m^*\dot{m} - m\dot{m}^* + h^*\dot{h} - h\dot{h}^* + V^*\dot{V} - V\dot{V}^*\right)$$
$$-\sum_{j,k} z_k^* A_{jk} z_j + \beta_n(V_m)nn^* + \alpha_n(V_m)(nn^* - n - n^*)$$
$$+ \beta_m(V_m)mm^* + \alpha_m(V_m)(mm^* - m - m^*) + \beta_h(V_m)hh^*$$
$$+ \alpha_h(V_m)(hh^* - h - h^*) + \frac{1}{\tau}VV^* - \frac{1}{\tau}V_L(V + V^*), \quad [47]$$

and we can use Eq [46] to show that the corresponding Hamiltonian reads:

$$\mathcal{H} = \sum_{j,k} z_k^* A_{jk} z_j - \beta_n(V_m)nn^* + \alpha_n(V_m)(n + n^* - nn^*) - \beta_m(V_m)mm^*$$
$$+ \alpha_m(V_m)(m + m^* - mm^*) - \beta_h(V_m)hh^*$$
$$+ \alpha_h(V_m)(h + h^* - hh^*) + \frac{1}{\tau}(V_L(V + V^*) - VV^*), \quad [48]$$

which we now differentiate in time as follows:

$$\dot{\mathcal{H}} = \sum_{j,k}\dot{z}_k^* A_{jk} z_j + \sum_{j,k} z_k^* A_{jk}\dot{z}_j - \beta_n(V_m)\dot{n}n^* - \beta_n(V_m)n\dot{n}^* - \beta_m(V_m)\dot{m}m^*$$
$$- \beta_m(V_m)m\dot{m}^* - \beta_h(V_m)\dot{h}h^* - \beta_h(V_m)h\dot{h}^*$$
$$+ \alpha_n(V_m)(\dot{n} + \dot{n}^* - \dot{n}n^* - n\dot{n}^*) + \alpha_m(V_m)(\dot{m} + \dot{m}^* - \dot{m}m^* - m\dot{m}^*)$$
$$+ \alpha_h(V_m)(\dot{h} + \dot{h}^* - \dot{h}h^* - h\dot{h}^*) + \frac{1}{\tau}(V_L(\dot{V} + \dot{V}^*) - \dot{V}V^* - V\dot{V}^*), \quad [49]$$

which, using Eqs [23] through [25], and dropping the subscripts, reads:

$$\dot{\mathcal{H}} = i(\dot{z}^*\dot{z} + \dot{n}^*\dot{n} + \dot{m}^*\dot{m} + \dot{h}^*\dot{h} + \dot{V}^*\dot{V} - \dot{z}\dot{z}^* - \dot{n}\dot{n}^* - \dot{m}\dot{m}^* - \dot{h}\dot{h}^* - \dot{V}\dot{V}^*) = 0. \quad [50]$$

Therefore, the energy does not change in time – i.e. it is conserved by virtue of time translational invariance in the complex oscillatory forms of the underlying models.

**Discussion**

The complex oscillatory forms of the models considered in Eqs [23] through [25] are fundamentally different to their original forms shown in Eqs [15] through [17]. This is due to the introduction of the imaginary unit $i$, which we showed is a necessary modification if we are to cast the equations of motion in the form of a Lagrangian. If we consider a generic first order differential equation written in terms of a complex dependent variable:

$$i\dot{z} = z \qquad [51]$$

we can re-write this by splitting the complex variable $z$ into its real and imaginary components according to $z = x + iy$, which means that:

$$i(\dot{x} + i\dot{y}) = x + iy \qquad [52]$$

where we can equate the real and imaginary components to write the following two equations:

$$\dot{y} = -x$$
$$\dot{x} = y \qquad [53]$$

or equivalently:

$$\ddot{x} = -x \qquad [54]$$

i.e. we see that the complex first order equation [51] is equivalent to the real second order equation [54]. It should therefore, in this sense, come as no surprise that we can cast a Lagrangian of models such as those in Eq [51], as these are just uncoupled harmonic oscillator descriptions as in Eq [54], which are readily described by Lagrangian mechanics. One advantage of working with complex variables is that the adjacency or coupling matrix A can be transformed into a symmetric form, where the real parts describe dissipation and the complex parts describe oscillatory or solenoidal dynamics (Friston et al., 2014, Friston et al., 2020) that underwrite rhythms in the brain (Buzsaki et al., 2013). More generally, the ability to work with non-dissipative solenoidal dynamics means that one can eschew detailed balance and characterise (neuronal) systems in terms of their nonequilibrium steady states (Friston, 2019, Yan et al., 2013, Kwon and Ao, 2011).

The imaginary unit $i$ famously appears in the Schrödinger equation $i\hbar\dot{\psi} = H\psi$, which we see possesses the same mathematical structure as the computational neuroscience models considered here: i.e. first order in time and linearly proportional to the dependent variable. As with the Schrödinger equation, this imaginary unit possesses a significance beyond facilitating Lagrangian formulations. To begin with, as shown in in Eqs [51] through [54], the imaginary unit renders the solutions of the models oscillatory. This oscillatory behaviour is one of the reasons Schrödinger introduced the imaginary unit in the context of quantum systems displaying wave-particle duality. The systems with which we deal in computational neuroscience are clearly different to those in quantum mechanics—we do not expect to see wave-particle duality in the brain. However, we have an equally valid reason for introducing the imaginary unit $i$, due to the fact that a neural region—even when unperturbed—often expresses intrinsic oscillatory behaviour (Buzsaki, 2006). As such, there is a case to be made that the modified form of the DCM neuronal state equation [23] is in fact more biologically plausible than the original in Eq [15]. However, in the case of the HH and LIF models it is not obvious whether this oscillatory form is appropriate in the description of cellular-level dynamics. It is possible in this regard that the use of the oscillatory forms of the HH and LIF models could be applicable to specific situations in which there is known oscillatory behaviour at microscopic scales (Li and Rinzel, 1994, Rossoni et al., 2005, Rubin and Wechselberger, 2008, Shorten and Wall, 2000).

The other reason Schrödinger introduced the imaginary unit in his equation is that it allows for the conservation of probability during the evolution of the wavefunction $\psi$. In quantum mechanics, the probability of observing an event is proportional to the squared modulus of the wavefunction:

$$\psi(t)\psi^*(t) \sim e^{i\omega t}e^{-i\omega t} = 1 \qquad [55]$$

i.e. we obtain unit probability and hence unitarity is preserved.

On the other hand, we can consider the consequences of calculating a probability in this way when the imaginary unit $i$ is not present:

$$\psi(t)\psi^*(t) \sim e^{\omega t} e^{\omega t} = e^{2\omega t} \qquad [56]$$

i.e. the probability is time dependent and is therefore not conserved.

The preservation of unitarity has potentially interesting consequences with regard to neural systems in the event that we are modelling probabilistic quantities, e.g. the chance of observing neural activation beyond a certain threshold. In this case, the presence of the imaginary unit $i$ in the modified equations of motion [23] through [25] imply a conservation of neuronal firing or depolarisation in the system, which could plausibly be maintained by the balance between excitation and and inhibition (Isaacson and Scanziani, 2011, Xue et al., 2014).

It should be noted that we are unlikely to be able to apply the principle of stationary action to neural systems in the same way as we would for a classical system, such as when calculating the trajectory of a cannonball. Instead, any application to neural systems is likely to be set in a statistical framework, in which we make predictions as to what the most probable evolution of the system will be. This may have a certain overlap with the path integral formulation of quantum mechanics (Feynman et al., 1948), in which e.g. an electron travels via an infinite number of paths between two points, each associated with a different value of the action. These paths are then summed in order to obtain the probability amplitude, the square of which yields the probability of the transition. In the case of neural systems, we may need to deal with a similar calculation as for the electron – i.e. considering different values of the action for the many ways in which the system can evolve.

In summary, we have demonstrated that it is possible to cast complex oscillatory versions of the dynamic causal modelling (DCM) neuronal state equation, Hodgkin-Huxley (HH) model, and leaky integrate-and-fire (LIF) model in the form of a single Lagrangian. We are therefore now in the position, using Eqs [5], [6], and [26], to express the evolution of neural systems described by any of these three models in terms of the principle of stationary action as follows:

$$\delta \int_{t_1}^{t_2} \left[ \frac{i}{2}\left( \sum_j (z_j^* \dot{z}_j - z_j \dot{z}_j^*) + n^* \dot{n} - n\dot{n}^* + m^* \dot{m} - m\dot{m}^* + h^* \dot{h} - h\dot{h}^* + V^* \dot{V} - V\dot{V}^* \right) \right.$$
$$- \sum_{j,k} z_k^* A_{jk} z_j - \sum_j \left( \sum_k C_k v_k + \omega_j^{(z)} \right)(z_j + z_j^*) + \beta_n(V_m) nn^*$$
$$+ \alpha_n(V_m)(nn^* - n - n^*) + \beta_m(V_m) mm^* + \alpha_m(V_m)(mm^* - m - m^*)$$
$$+ \beta_h(V_m) hh^* + \alpha_h(V_m)(hh^* - h - h^*) + \frac{1}{\tau} VV^*$$
$$\left. - \frac{1}{\tau}(RI + V_L)(V + V^*) \right] dt = 0 \qquad [57]$$

This formalism allows for three examples of equations of motion in computational neuroscience—hereto considered separately—to be cast in a unified form in terms of the principle of stationary action. Importantly, this expression entails a unification across scales, as the models contained within this expression are used to describe the ways in which neural systems evolve from microscopic (HH & LIF) through to macroscopic (DCM) scales. We are also able to define analogues of quantities such as momentum (see Eq [45]) and energy (see Eq [46]) within a neural system. These quantities are of course difficult to interpret at this stage, as they are yet to be mapped onto biological mechanisms: this remains a subject for future research. The focus of this work is to create the first steps in allowing computational neuroscience to be written in the same language as a vast amount of the physical sciences, in the hope that this common framework will yield new insight into hereto disparate analyses of cross-scale neural dynamics.

ABBOTT, L. F. 1999. Lapicque's introduction of the integrate-and-fire model neuron (1907). *Brain research bulletin,* 50**,** 303-304.

BREAKSPEAR, M. 2017. Dynamic models of large-scale brain activity. *Nature neuroscience,* 20**,** 340-352.

BUZSAKI, G. 2006. *Rhythms of the Brain*, Oxford University Press.

BUZSAKI, G., LOGOTHETIS, N. & SINGER, W. 2013. Scaling brain size, keeping timing: evolutionary preservation of brain rhythms. *Neuron,* 80**,** 751-64.

COTTINGHAM, W. N. & GREENWOOD, D. A. 2007. *An introduction to the standard model of particle physics*, Cambridge university press.

DECO, G., JIRSA, V. K., ROBINSON, P. A., BREAKSPEAR, M. & FRISTON, K. J. 2008. The Dynamic Brain: From Spiking Neurons to Neural Masses and Cortical Fields. *Plos Computational Biology,* 4.

FEYNMAN, R., DEUTSCH, A. J., ECKART, C., COWAN, R. & DIEKE, G. 1948. Space-Time Approach to Non-Relativistic Quantum Mechanics. *Space,* 20.

FRISTON, K. J., HARRISON, L. & PENNY, W. 2003. Dynamic causal modelling. *Neuroimage,* 19**,** 1273-302.

HODGKIN, A. L. & HUXLEY, A. F. 1952. A quantitative description of membrane current and its application to conduction and excitation in nerve. *The Journal of physiology,* 117**,** 500.

ISAACSON, J. S. & SCANZIANI, M. 2011. How inhibition shapes cortical activity. *Neuron,* 72**,** 231-43.

LANDAU, L. D. 2013. *The classical theory of fields*, Elsevier.

LANDAU, L. D., LIFSHITZ, E.M. 1976. *Mechanics (third edition), Vol. 1 of Course of Theoretical Physics,* Oxford, Pergamon Press.

LI, B., DAUNIZEAU, J., STEPHAN, K. E., PENNY, W., HU, D. & FRISTON, K. 2011. Generalised filtering and stochastic DCM for fMRI. *Neuroimage,* 58**,** 442-57.

LI, Y.-X. & RINZEL, J. 1994. Equations for InsP3 receptor-mediated [Ca2+] i oscillations derived from a detailed kinetic model: a Hodgkin-Huxley like formalism. *Journal of theoretical Biology,* 166**,** 461-473.

MORSE, P. M. & FESHBACH, H. 1953. *The Variational Integral and the Euler Equations,* New York, McGraw-Hill.

NOETHER, E. 1918. Invariante Variationsprobleme. *Nachrichten von der Königlichen Gesellschaft der Wissenschaften zu Göttingen. Mathematisch-physikalische Klasse*, 235-257.

ROSSONI, E., CHEN, Y., DING, M. & FENG, J. 2005. Stability of synchronous oscillations in a system of Hodgkin-Huxley neurons with delayed diffusive and pulsed coupling. *Physical Review E,* 71**,** 061904.

RUBIN, J. & WECHSELBERGER, M. 2008. The selection of mixed-mode oscillations in a Hodgkin-Huxley model with multiple timescales. *Chaos: An Interdisciplinary Journal of Nonlinear Science,* 18**,** 015105.

SHANKAR, R. 2012. *Principles of quantum mechanics*, Springer Science & Business Media.

SHORTEN, P. R. & WALL, D. J. 2000. A Hodgkin–Huxley model exhibiting bursting oscillations. *Bulletin of mathematical biology,* 62**,** 695-715.

STEPHAN, K. E., KASPER, L., HARRISON, L. M., DAUNIZEAU, J., DEN OUDEN, H. E., BREAKSPEAR, M. & FRISTON, K. J. 2008. Nonlinear dynamic causal models for fMRI. *Neuroimage,* 42**,** 649-62.

XUE, M., ATALLAH, B. V. & SCANZIANI, M. 2014. Equalizing excitation-inhibition ratios across visual cortical neurons. *Nature,* 511**,** 596-600.

**Author contributions:** All authors conceived of the analysis and wrote the paper.

**Acknowledgements:** We thank W.M.C. Foulkes, Imperial College London for his invaluable help; E.D.F. was supported by a King's College London Prize Fellowship; K.J.F. was funded by a Wellcome Principal Research Fellowship (Ref: 088130/Z/09/Z); R.J.M was funded by the Wellcome/EPSRC Centre for Medical Engineering (Ref: WT 203148/Z/16/Z); R.L was funded by the MRC (Ref: MR/R005370/1). The authors would also like to acknowledge support from the Data to Early Diagnosis and Precision Medicine Industrial Strategy Challenge Fund, UK Research and Innovation (UKRI), the National Institute for Health Research (NIHR), the Biomedical Research Centre at South London, the Maudsley NHS Foundation Trust, and King's College London.

**Competing interests:** The authors declare no competing interests.